\documentstyle[amssymb,12pt]{article}

\begin{document}

\title{Accelerated Charge Kerr-Schild Metrics in\\ $D$- Dimensions}
\author{ Metin G{\" u}rses\\
{\small Department of Mathematics, Faculty of Sciences}\\
{\small Bilkent University, 06533 Ankara - Turkey}\\
\\
{\" O}zg{\" u}r Sar{\i}o\u{g}lu  \\
{\small Department of Physics, Faculty of Arts and  Sciences}\\
{\small Middle East Technical University, 06531 Ankara Turkey}}
\begin{titlepage}
\maketitle
\begin{abstract}
We consider the $D$ dimensional Einstein Maxwell theory with
a null fluid in the Kerr-Schild Geometry. We obtain a complete set of 
differential conditions that are necessary for finding solutions.
We examine the case of vanishing  pressure and cosmological
constant in detail. For this specific case, we give the metric,
the electromagnetic vector potential and the fluid energy density.
This is, in fact, the generalization of the well known Bonnor-Vaidya
solution to arbitrary $D$ dimensions. We show that due to the
acceleration of charged sources, there is an energy flux in $D \ge 4$ 
dimensions and we give the explicit form of  this energy flux formula.

\end{abstract}
\end{titlepage}

\section{Introduction}

Radiation  and hence energy loss  due to the
acceleration of an electron is a well known phenomenon in classical
electromagnetism. An exact solution describing this phenomenon in
General Relativity is due to Bonnor and Vaidya \cite{bv1}, where
the metric is in Kerr-Schild form \cite{ks}, \cite{gg1}, \cite{crm}.
An acceleration parameter has  also been considered in Robinson-Trautman
metrics \cite{rt1}-\cite{rt3}.
The energy loss formula turns out to be exactly the same as the one
obtained from classical electromagnetism.
When the acceleration vanishes, Bonnor-Vaidya metrics reduce to the 
Reissner-Nordstr{\" o}m (RN) metric.

In this work our main motivation is to generalize Bonnor-Vaidya and Photon  
Rocket solutions of $D=4$ General Relativity. For this purpose 
we consider the $D$-dimensional Kerr-Schild
metric with an appropriate vector potential and a fluid velocity
vector, and we derive a complete set of conditions for the Einstein 
Maxwell theory with a null perfect fluid. We find the 
expressions for the pressure and the mass density of the fluid. 
We classify our solutions under some certain assumptions.
The field equations are highly complex and with some nontrivial
equation of state it is quite difficult to solve the corresponding
equations. To this end we assume vanishing pressure and  cosmological constant.
 We give the complete solution for  arbitrary dimension $D$.
This generalizes the Bonnor-Vaidya solution. We obtain
the energy flux formula depending on the dimension $D$.  

Our conventions are similar to the conventions of Hawking-Ellis \cite{hwk}.
This means that the Riemann tensor $R^{\alpha}\,\,_{\beta \mu \nu}$,
 Ricci tensor $R_{\alpha \beta}$, Ricci scalar $R_{s}$ and the Einstein
tensor $G_{\mu \nu}$ are defined by

\begin{eqnarray}
R^{\alpha}\,\,_{\beta \mu \nu}=
\Gamma^{\alpha}\,_{\beta \nu, \mu}-\Gamma^{\alpha}\,_
{\beta \mu, \nu}+\Gamma^{\alpha}\,_{\mu \gamma}\,
 \Gamma^{\gamma}\,_{\beta \nu}-
\Gamma^{\alpha}\,_{\nu \gamma}\, \Gamma^{\gamma}\,_{\beta \mu},\\
R_{\alpha \beta}=R^{\gamma}\,\,_{\alpha \gamma \beta},~~~ R_{s}=
R^{\alpha}\,_{\alpha},~~G_{\alpha \beta}=R_{\alpha \beta}-{1 \over 2}
 g_{\alpha \beta}\,R_{s}
\end{eqnarray}

\noindent
Here $g_{\mu \nu}$ is the $D$ dimensional metric tensor with 
signature $(-, +, +, \cdots, +)$. The source of the Einstein equations
is composed of the electromagnetic field with the vector potential
field $A_{\mu}$  and a perfect fluid with velocity vector field
$u^{\mu}$, energy density $\rho$ and pressure $p$. Their energy momentum
 tensors are respectively given by

\begin{eqnarray}
T^{e}_{\mu \nu}&=&F_{\mu }\,^{\alpha}\, F_{\nu \alpha}-{1 \over 4} 
F^2 g_{\mu \nu},~~ F^2 \equiv F^{\alpha \beta}\,F_{\alpha \beta},\\
T^{f}_{\mu \nu}&=& (p+\rho)\, u_{\mu}\, u_{\nu}+ p g_{\mu \nu}
\end{eqnarray}

\noindent
Then the Einstein equations are given by

\begin{eqnarray}
&&G_{\mu \nu}=\kappa T_{\mu \nu}=\kappa \, (T^{e}_{\mu \nu}+ T^{f}_{\mu \nu})
+\Lambda g_{\mu \nu}, \label{ein1}\\
&&(p+\rho) \, u^{\nu}\, u^{\mu}\,_{ ; \nu}=-u^{\nu}\,(\rho \, u^{\mu})_{; \nu}+
p_{,\nu}\,(g^{\mu \nu}+u^{\mu} \, u^{\nu})+F^{\mu}\,_{\nu}\, J^{\nu},
\label{ein2}\\
&&F^{\mu \nu}\,\,_{;\nu}= J^{\mu} \label{ein3}
\end{eqnarray}

\noindent
In the next section, we develop the kinematics of a curve $C$ in the $D$-
dimensional Minkowski manifold ${\bf M_{D}}$. We construct solutions of the
electromagnetic vector field due to the acceleration of charged particles
in four and six dimensions.
We then find the energy flux due to  acceleration. 
In Section 3, we give a detailed study of Kerr-Schild Geometry under certain
assumptions. 
In Section 4, we find
the solution of the $D$- dimensional Einstein Maxwell field equations with 
a null fluid. We also obtain the generalization of Bonnor-Vaidya metrics 
to $D$- dimensions. We derive the energy flux formula depending on the 
dimension $D$ and discuss its finiteness in Section 5 and state our 
conclusion in Section 6. Finally , for convenience to the reader,
in an Appendix we give some well-known formulae that are needed in the text.

\section{Radiation due to Acceleration: Maxwell Theory}

Let $z^{\mu}(\tau)$ describe a smooth curve $C$ defined by 
$z: I \subset {\bf R}
\rightarrow {\bf M_{D}}$. Here $\tau \in I$, $I$ is an interval on the 
real line and 
${\bf M_{D}}$ is the $D$- dimensional Minkowski manifold. From an arbitrary
point $x^{\mu}$ outside the curve, there are two null lines intersecting the
curve $C$. These points are called the retarded and advanced times. Let
$\Omega$ be the distance between the points $x^{\mu}$ and $z^{\mu}(\tau)$,
so by definition it is given  by

\begin{equation}
\Omega=\eta_{\mu \nu}\,(x^{\mu}-z^{\mu}(\tau))\,(x^{\nu}-z^{\nu}(\tau)),
\label{dist}
\end{equation}

\noindent
Hence $\Omega=0$ for two values of $\tau$ for a non-space-like curve.
Let us call these as, $\tau_{0}$ (retarded) and $\tau_{1}$ (advanced). We shall
focus ourselves to the retarded case only. The main reason for this is that the
Green's function for the vector potential chooses this point on the
curve $C$ (\cite{brt}, \cite{jack}).

If we differentiate $\Omega$ with respect to $x^{\mu}$ and let
$\tau=\tau_{0}$, then we get

\begin{equation}
\lambda_{\mu} \equiv \tau_{,\mu}={x_{\mu}-z_{\mu}(\tau_{0}) \over R}, 
\label{lam}
\end{equation}

\noindent
where $\lambda_{\mu}$ is a null vector and $R$ is the retarded distance
defined by

\begin{equation}
R \equiv  \dot{z^{\mu}}(\tau_{0})\,(x_{\mu}-z_{\mu}(\tau_{0})).
\end{equation}

\noindent
Here a dot over a letter denotes differentiation with respect to $\tau_{0}$.
We now list some properties of $R$ and $\lambda_{\mu}$.

\begin{eqnarray}
\lambda_{\mu, \nu}&=&{1 \over R}\,[\eta_{\mu \nu}-\dot{z}_{\mu}\, \lambda_{\nu}
-\dot{z}_{\nu}\, \lambda_{\mu}-(A-\epsilon)\, \lambda_{\mu} \lambda_{\nu}],\\
R_{,\mu}&=&(A-\epsilon)\, \lambda_{\mu}+ \dot{z}_{\mu}
\end{eqnarray}

\noindent
where 

\begin{equation}
A=\ddot{z}^{\mu}\,(x_{\mu}-z_{\mu}),~~~ \dot{z}^{\mu}\, \dot{z}_{\mu}=
\epsilon= 0, \pm 1
\end{equation}

\noindent
Here $\epsilon =-1$ for time-like curves and vanishes for null curves.
Furthermore we have

\begin{equation}
\lambda_{\mu}\, \dot{z}^{\mu}=1,~~~ \lambda^{\mu}\, R_{, \mu}=1
\end{equation}

\noindent
Let $ a={A \over R}$. Then it is easy to prove that 

\begin{equation}
a_{,\mu}\, \lambda^{\mu}=0 \label{cond}
\end{equation}

\noindent
Similar to $a$, we have other scalars satisfying the same property (\ref{cond})
obeyed by $a$.

\vspace{0.3cm}

\noindent
{\bf Lemma 1}.\, {\it Let 

\begin{equation}
a_{k}=\lambda_{\mu}\, {d^k \, \ddot{z^{\mu}} \over d\tau_{0}^k},
~~~k=1,2,\cdots ,n
\label{ak}
\end{equation}

\noindent
then 

\begin{equation}
a_{k, \alpha}\, \lambda^{\alpha}=0, \label{con01}
\end{equation}

\noindent
for all $k$. Furthermore if $A_{k}=R \, a_{k}$ is constant for a fixed
$k$ at all points then
$A_{i}=0$ for all $i \ge k$ and $d^m \ddot{z}^{\mu}/d\tau_{0}^m=0$ for all
$m \ge k$.  Here $n$ is an arbitrary positive integer.
}

\vspace{0.3cm}

\noindent
{\bf Proof}.\, The proof of this Lemma depends on the following
formula for the derivative of $A_{k}$

\begin{equation}
A_{k, \alpha}={d^{k} \ddot{z}_{\alpha} \over d\tau_{0}^{k}}+
[A_{k+1}-(d^k \ddot{z}^{\beta}/d\tau_{0}^k)\, \dot{z}_{\beta}]\,
\lambda_{\alpha}. \label{ak1}
\end{equation}

\noindent
In the sequel, for the sake of simplicity,
we shall use $\tau$ instead of $\tau_{0}$.
First part of the Lemma can be proved by contracting the above formula 
(\ref{ak1}) by $\lambda^{\alpha}$. One obtains 

\begin{equation}
\lambda^{\alpha}\, A_{k, \alpha}=a_{k}. \label{ak2}
\end{equation}

\noindent
This implies that $\lambda^{\alpha}\, a_{k,\alpha}=0$.
For the second part of the Lemma contracting the above formula (\ref{ak1}) by 
$\dot{z}^{\alpha}$ one obtains 

\begin{equation}
A_{k, \, \alpha}\, \dot{z}^{\alpha}=A_{k+1}, \label{ak3}
\end{equation}

\noindent
which implies that $A_{m}=0$ for $ m > k$. From (\ref{ak2}) we have also
$A_{k}=0$. With these results (\ref{ak1}) reduces to

\begin{equation}
{d^k \ddot{z}_{\alpha} \over d\tau^k}= \mu \, \lambda_{\alpha},~~~
\mu=(d^k \ddot{z}^{\beta}/d\tau^k)\, 
\dot{z}_{\beta}. \label{ak4}
\end{equation}

\noindent
Differentiating this equation with respect to $x^{\beta}$ and contracting
the resulting equation by $\lambda^{\beta}$ we get $\mu=0$. This completes
the proof of the Lemma.

\vspace{0.3cm}

In (\ref{ak}) $n$ depends on the dimension $D$ of the manifold ${\bf M_{D}}$. 
The scalars
 $(a,\,a_{k})$, are related to the  curvature scalars of the curve $C$ 
in ${\bf M_{D}}$. The
number of such scalars is $D-1$ \cite{spv}, \cite{guv}. Hence we let $n=D-1$. 

Before moving on to the  main subject of this work, i.e. examining the 
Einstein-Maxwell theory within the Kerr-Schild geometry, we briefly study 
the Maxwell theory in flat (even) $D$-dimensional Minkowski space. This
should at least serve the reader as a reminder of the well-known basics
regarding the usual $D=4$ Maxwell theory. 

By  using the above curve $C$ and its kinematics, we can construct divergence
free (Lorentz gauge) vector fields $A_{\alpha}$ satisfying the 
wave equation $ \eta^{\mu \nu}\, \partial_{\mu}\, \partial_{\nu}\, A_{\alpha}
=0$ outside 
the curve $C$ in any even dimension $D$. For instance in the cases 
$D=4$ and $D=6$ we have respectively \cite{go1}, \cite{go2}

\begin{equation}
A_{\mu}= \left \{ \begin{array}{ll}{e \over 4\pi}\,{\dot{z}_{\mu} \over R},
& (\mbox{$D = 4$})\\
{e \over 4\pi^2}\,[{\ddot{z}_{\mu}-a\, \dot{z}_{\mu} \over R^2} +
 \epsilon{ \dot{z}_{\mu} \over R^3}] & (\mbox{$D = 6$})
\end{array} 
\right. \label{a66}
\end{equation}

\noindent
These vector fields represent respectively the
electromagnetic vector potentials of an accelerated charge in four and 
six dimensional Maxwell theory of electromagnetism. The flux of
electromagnetic energy is then given by \cite{brt}

\begin{equation}
dE=-\int_{S}\, \dot{z}_{\mu}\,T^{\mu \nu} dS_{\nu}
\end{equation}

\noindent
where $T_{\mu \nu}=F_{\mu \alpha}\, F_{\nu}\,^{\alpha}-{1 \over 4} F^2 
\eta_{\mu \nu}$ is the Maxwell energy momentum tensor, $F_{\mu \nu}=
A_{\nu,\mu}-A_{\mu,\nu}$ is the  electromagnetic field tensor
 and $F^2=F^{\alpha \beta}\,F_{\alpha \beta}$.
 The surface element $dS_{\mu}$ on $S$ is given by

\begin{equation}
dS_{\mu}=n_{\mu} R d\tau\, d\Omega
\end{equation}

\noindent
where $n_{\nu}$ is orthogonal to the velocity vector field $\dot{z}_{\mu}$
which is defined through

\begin{equation}
\lambda_{\mu}=\epsilon \dot{z}_{\mu}+\epsilon_{1}\,{n_{\mu} \over R},
~~~ n^{\mu}\, n_{\mu}=-\epsilon R^2
\end{equation}

\noindent
Here $\epsilon_{1}=\pm 1$.
For the remaining part of this section we shall assume $\epsilon=-1$
($C$ is a time-like curve).
One can consider $S$ in the rest frame as a sphere of radius $R$. Here
$d\Omega$ is the solid angle. Letting $dE/d\tau=N_{e}$ \cite{bv1}, 
then we have

\begin{equation}
N_{e}=-\int_{S}\, \dot{z}_{\mu}\,T^{\mu \nu}\, n_{\nu} \, R d\Omega
\end{equation}

\noindent
At very large values of $R$ we get for $D=4$

\begin{eqnarray}
N_{e}&=& (e / 4\pi)^2\, \epsilon_{1}\, \int\, 
(- \ddot{z}_{\mu}\, \ddot{z}^{\mu}+a^2)\,d\Omega\, \\
&=&- \epsilon_{1}\,(e/4 \pi)^2\, (\ddot{z}^{\mu}\, \ddot{z}_{\mu}) \int\,
(1-\cos^2\theta) \sin \theta\, d\theta d\phi,\\
&=&- \epsilon_{1}\,(e^2/4\pi) \,{2 \over 3}\, (\ddot{z}^{\mu}\, \ddot{z}_{\mu})
\label{los1}
\end{eqnarray}

\noindent
and for $D=6$ at very large values of $R$ we have

\begin{eqnarray}
N_{e}&=&-\int_{S}\, \dot{z}_{\mu}\, T^{\mu \nu}\, n_{\nu} R^3\, 
d \Omega,\\
&=&-({e \over 4\pi^2})^2 \epsilon_{1}\, \int\, \xi_{\mu} \xi^{\mu} d\Omega
\end{eqnarray}

\noindent
where 

\begin{equation}
\xi^{\mu}={d^3\, z^{\mu} \over d\tau^3}-3 a {d^2 \, z^{\mu} \over d\tau^2}
+(-a_{1}+3 a^2)\, {d\, z^{\mu} \over d\tau}
\end{equation}

\noindent
so that $\lambda_{\mu}\, \xi^{\mu}=0$. For a charge $e$ with 
acceleration $|\ddot{z}_{\mu}|=\kappa_{1}$ we have (for $D=6$ case)

\begin{equation}
N_{e}=-({e \over 4\pi^2})^2\, {32 \pi^2 \epsilon_{1} \over 15}
(\dot{\kappa}_{1}^2+\kappa_{1}^2 \kappa_{2}^2 +{9 \over 7}\, \kappa_{1}^4)
\label{los2}
\end{equation}

\vspace{0.3cm}

\noindent
{\bf Remark 1}.\, To be compatible with the classical results \cite{brt},
\cite{jack} we take $\epsilon_{1}= -1$. We also conjecture that the sign 
of the energy flux will be the same for all even dimensions.

\vspace{0.3cm}

\section{Accelerated Kerr-Schild Metrics in $D$ - Dimensions}

We now consider the Einstein-Maxwell field equations with a perfect fluid
distribution in $D$-dimensions.
Here we make some assumptions. First of all, we assume that the metric of the 
$D$-dimensional space-time is the Kerr-Schild metric. Furthermore, we take
the null vector $\lambda_{\mu}$ in the metric as the same  null vector 
defined in (\ref{lam}). With these assumptions the Ricci tensor takes 
a special form.

\vspace{0.3cm}

\noindent
{\bf Proposition 2}.\, {\it Let $g_{\mu \nu }=\eta_{\mu \nu}-2V \lambda_{\mu}\,
\lambda_{\nu}$ and $\lambda_{\mu}$ be the null vector defined in 
(\ref{lam}) and
let $V$ be a differentiable function, then the Ricci tensor and the
Ricci scalar are given by

\begin{eqnarray}
R^{\alpha}\,_{\beta}&=&\zeta_{\beta}\, \lambda^{\alpha}+
\zeta^{\alpha}\, \lambda_{\beta}+r \delta^{\alpha}\,_{\beta}
+q \lambda_{\beta}\, \lambda^{\alpha},\\
R_{s}&=&-2 \lambda^{\alpha}\,K_{,\alpha}-4\theta K-{2V \over R^2}(D-2)(D-3),
\label{rics}
\end{eqnarray}

\noindent
where 

\begin{eqnarray}
r&=&{2(-D+3)\, V \over R^2}-{2K \over R},   \label{rrr}\\
q&=&\eta^{\alpha \beta}\, V_{, \alpha \beta}+\epsilon r+{2A \over R}(K+
\theta V)-{4 \over R}(\dot{z}^{\mu}\, V_{,\mu}+AK-\epsilon K), \label{rho}\\
\zeta_{\alpha}&=&-K_{,\alpha}-{D-4 \over R} V_{, \alpha}+{2V \over R^2} (D-3)\,
\dot{z}_{\alpha}, \label{zet}
\end{eqnarray}

\noindent
where $K \equiv \lambda^{\alpha}\, V_{,\alpha}$ and $\theta \equiv
\lambda^{\alpha}\,_{,\alpha}={D-2 \over R}$.
}

\vspace{0.3cm}

\noindent
Let us further assume that the electromagnetic vector potential $A_{\mu}$ is given by
$A_{\mu}=H\, \lambda_{\mu}$ where $H$ is a differentiable function. Let $p$
and $\rho$ be the pressure and the energy density of a perfect fluid distribution
with the velocity vector field $\lambda_{\mu}$. Then the difference tensor
${\cal T}^{\mu}\,_{\nu}=G^{\mu}\,_{\nu}-\kappa T^{\mu}\,_{\nu}$
is given by the following proposition:

\vspace{0.3cm}

\noindent
{\bf Proposition 3}.\, {\it Let $g_{\mu \nu}=\eta_{\mu \nu}-2V \lambda_{\mu}
 \lambda_{\nu}$, $A_{\mu}=H \lambda_{\mu}$, where $\lambda_{\mu}$ is given
in (\ref{lam}), $V$ and $H$ be   differentiable functions. Let $p$ and 
$\rho$ be the pressure and energy density of a perfect fluid with velocity
vector field $\lambda_{\mu}$. Then the difference tensor becomes

\begin{equation}
{\cal T}^{\alpha}\,_{\beta}=\lambda^{\alpha}\, W_{\beta}+
\lambda_{\beta}\, W^{\alpha}+{\cal P}\, \delta^{\alpha}\,_{\beta}+
{\cal Q}\, \lambda^{\alpha}\, \lambda_{\beta}
\end{equation}

\noindent
where

\begin{eqnarray}
{\cal P}&=&r-{1 \over 2}R_{s}-{1 \over 2}\kappa (\lambda^{\mu}\, H_{,\mu})^2
-(\kappa p+\Lambda),\\
{\cal Q}&=&q-\kappa (p+\rho)-\kappa (\eta^{\alpha \beta}\, 
H_{,\alpha}\, H_{,\beta}),\\
W_{\alpha}&=&\zeta_{\alpha}+\kappa (\lambda^{\mu} H_{,\mu}) H_{,\alpha} ,
\label{om01}
\end{eqnarray}

\noindent
where $\zeta_{\alpha}$, $r$, $R_{s}$ and $q$ are given in (\ref{zet}), 
(\ref{rrr}), (\ref{rics}) and (\ref{rho}) respectively.
}

\vspace{0.3cm}

\noindent
The vanishing of the difference tensor ${\cal T^{\alpha}\,_{\beta}}$ in 
Proposition 3 implies that ${\cal P}=0$ and $W_{\alpha}=-{{\cal Q} \over 2}\, 
\lambda_{\alpha}$. Hence
the following Corollary gives a set of equations that are equivalent to
the Einstein equations under the assumptions of Proposition 3.

\vspace{0.3cm}

\noindent
{\bf Corollary 4}.\,{\it With the assumptions of Proposition 3 , the Einstein
equations (\ref{ein1})  imply that

\begin{eqnarray}
\kappa p+ \Lambda &=& {1 \over 2} \kappa (\lambda^{\alpha}\, H_{, \alpha})^2
+{D-2 \over R}\,K+{(D-2)(D-3) \over R^2}V, \label{cor1}\\
\kappa (p+\rho)&=&q-\kappa \eta^{\alpha \beta}\, H_{,\alpha} H_{,\beta}+2 w 
\label{cor2}\\
W_{\alpha}&=&w \lambda_{\alpha} \label{cor3}
\end{eqnarray}

\noindent
where $w=W^{\alpha}\, \dot{z}_{\alpha}$ and $\Lambda$ is the cosmological 
constant. Here $q$ and $W_{\alpha}$ are respectively given in (\ref{rho})
and (\ref{om01}).
}

\vspace{0.3cm}

\noindent
We shall now assume that the functions $V$ and $H$ depend on $R$ and on some
$R$-independent functions $c_{i}$, ($i=1,2,\cdots$) such that

\begin{equation}
c_{i , \alpha}\, \lambda^{\alpha}=0, \label{con1}
\end{equation}

\noindent
for all $i$. It is clear that due to the property (\ref{cond}) and
(\ref{con01}) of $a_{k}$'s, 
all of these functions ($c_{i}$) are functions of the scalars $a$ and $a_{k}$,
$(k=1,2,\cdots, D-1)$. 

Notice that there are in fact $D+2$ equations contained in 
(\ref{cor1})-(\ref{cor3}).
In particular using the vector equation (\ref{cor3}) we can produce 
other scalar equations by contracting it with the vectors $\lambda^{\alpha}$
and $\dot{z}^{\alpha}$. We summarize the results in the following 
Proposition.

\vspace{0.3cm}

\noindent
{\bf Proposition 5}.\, {\it Let $V$ and $H$ depend on $R$ and functions
$c_{i}$ , $(i=1,2, \cdots)$ that satisfy (\ref{con1}), then the Einstein 
equations given in Proposition 4 reduce to the following set of equations

\begin{eqnarray}
&&\kappa p+\Lambda ={1 \over 2} V^{\prime \prime}+{3D-8 \over 2R} V^{\prime}
+{(D-3)^2 \over R^2} V, \label{pres}\\
&&\kappa (H^{\prime})^2= V^{\prime \prime}+{D-4 \over R} V^{\prime}-
{2V \over R^2}(D-3), \label{denk}\\
&&\kappa (p+\rho)=q-\kappa \eta^{\alpha \beta}\, H_{,\alpha} H_{,\beta}+
\nonumber\\
&&2\,[{2(A-\epsilon)(D-3)V \over R^2}-\sum_{i=1}(w_{i}\,
c_{i, \alpha}\, \dot{z}^{\alpha})], \label{rho1}\\
&&\sum_{i=1}\, w_{i}\,c_{i, \alpha}=[\sum_{i=1}\, (w_{i}\, c_{i, \beta}
\,\dot{z}^{\beta})]\, \lambda_{\alpha}, \label{cler}
\end{eqnarray}

\noindent
where

\begin{equation}
w_{i}=V^{\prime}_{,c_{i}}+{D-4 \over R}\, V_{,c_{i}}-
\kappa H^{\prime}\, H_{,c_{i}} ,
\end{equation}

\noindent
and  prime over a letter denotes partial differentiation with respect to
$R$. Eq. (\ref{ein2}) is satisfied identically with the electromagnetic current
vector

\begin{eqnarray}
J_{\mu}=  \sum_{i=1}\, \left [{1 \over R}(4-D) H_{,c_{i}}-
(H^{\prime})_{,c_{i}} \right ]\, 
c_{i,\mu}+[-H^{\prime \prime}+{1 \over R}(2-D) H^{\prime}]\,   
\dot{z}_{\mu} \nonumber \\
+ \left [\sum_{i=1}\, \left (2 (H^{\prime})_{,c_{i}} (c_{i, \alpha} 
\dot{z}^{\alpha})+
H_{,c_{i}} (c_{i,\alpha}\,^{,\alpha})-{2 \over R} H_{,c_{i}} (c_{i, \alpha}\,
\dot{z}^{\alpha}) \right )+A H^{\prime \prime } \right ]\, \lambda_{\mu}
\end{eqnarray}
}

\vspace{0.3cm}

\noindent
Notice that (\ref{denk}) is obtained by contracting (\ref{cor3}) with the 
vector $\lambda ^{\alpha}$.
The above equations can be described as follows. The equations (\ref{pres})
and (\ref{rho1}) define the pressure and mass density of the perfect fluid
distribution with null velocity $\lambda_{\mu}$. Eq.(\ref{denk}) gives a 
relation between the electromagnetic and gravitational potentials $H$ and $V$.
Since this relation is quite simple, given one of them one can easily 
solve the other one. Equation (\ref{cler}) implies that there are some
functions $c_{i}$\,  ($i=1,2,\cdots$)  where this equation is satisfied. 
The functions $c_{i}$  \, ($i=1,2, \cdots$) arise as integration constants
(with respect to the variable $R$) while determining the $R$ dependence
of the functions $V$ and $H$. 
Assuming the existence of such $c_{i}$'s the above
equations give the most general solution of the Einstein-Maxwell field
equations with a null perfect fluid distribution under the assumptions of 
Proposition 3.

\vspace{0.3cm}

\section{Null-Dust Solutions in $D$-Dimensions}

In this Section we  give a class of new exact solutions 
in the Kerr-Schild geometry.
Assuming that the null fluid has no pressure, the cosmological constant
vanishes and  using  Proposition 5, we have the following result:

\noindent
{\bf Theorem 6}.\, {\it Let $p=\Lambda=0$. Then

\begin{eqnarray}
V&=& \left \{\begin{array}{ll}
{\kappa e^2\,(D-3) \over 2(D-2)} R^{-2D+6}+m R^{-D+3} & (\mbox{$ D \ge 4$})\\
-{\kappa \over 2}e^2 \, ln R+m  &( \mbox{$D=3$})
\end{array} 
\right. \label{e50}\\
H&=& \left \{\begin{array}{ll}
c+\epsilon \,e\, R^{-D+3}  & ( \mbox{$D \ge 4$})\\
c+\epsilon e\, ln R  & ( \mbox{$D=3$})
\end{array} 
\right. \label{e51}
\end{eqnarray}

\noindent
where for $D > 3$

\begin{eqnarray}
\rho&=&-(c_{,\alpha}\, c^{,\alpha})
-\epsilon\,(3-D)\,e\, R^{3-D}\,(c^{, \alpha}\,_{, \alpha})-
\epsilon\,2(3-D)(2-D) e (c_{,\alpha}\,\dot{z}^{\alpha})\,R^{2-D} \nonumber \\
&&+{1 \over \kappa}\,a (2-D)(1-D) M R^{2-D} 
+(2-D)(3-D) a e^2 \,R^{5-2D} \nonumber \\
&&-\epsilon\, (2-D)(1-D)(3-D) a c e R^{2-D} 
+{1 \over \kappa} \dot{M} (2-D) R^{2-D} \nonumber \\
&&-\epsilon\,(3-D)(2-D) c \dot{e} R^{2-D} 
+(3-D) e \dot{e} R^{5-2D}, \label{ro00}\\
J_{\mu}&=&{1 \over R}(4-D) c_{, \mu}+[c^{,\alpha}\,_{,\alpha}-{2 \over R} 
(c_{,\alpha} \dot{z}^{\alpha}) \nonumber \\
&&+\epsilon (3-D) \dot{e} R^{2-D}+\epsilon (3-D)(2-D) e a R^{2-D}] 
\lambda_{\mu}
\end{eqnarray}

\noindent
and for $D=3$

\begin{eqnarray}
\rho&=&-(c_{,\alpha}\, c^{,\alpha})-\epsilon\, e (c_{,\alpha}\,^{,\alpha})
+\epsilon\,{2e \over R}\,(c_{,\alpha}\,
\dot{z}^{\alpha})-{a e^2 \over 2 R} \nonumber \\
&&+{2 M a \over \kappa R} 
- {a e^2 \over R}\, \ln R-
{\dot{M} \over \kappa R}+{e \dot{e} \over R}+\epsilon\,{c \dot{e} \over R}
\nonumber \\
&&+{e\dot{e} \over R} \ln R-2\epsilon\, ec {a \over R}, \label{ro11}\\
J_{\mu}&=&{1 \over R} c_{,\mu}+[c^{,\alpha}\,_{,\alpha}-{2 \over R} 
(c_{,\alpha} \dot{z}^{\alpha})+{\epsilon \over R} \dot{e}-\epsilon e
 {a \over R}] \lambda_{\mu}
\end{eqnarray}

\noindent
Here $M=m+\epsilon\,\kappa (3-D) e c$ for $D \ge 4$ and $M=m+{\kappa \over 2}\,e^2
+\epsilon\, \kappa e c$ for $D=3$.
In all cases $e$ is assumed to be a function of $\tau$ only but the functions
$m$ and $c$ which are related through the arbitrary function $M(\tau)$
 (depends on
$\tau$ only) do depend on the scalars $a$ and $a_{k}$, ($k \ge 1$). 
}

\vspace{0.3cm}

\noindent
Equation (\ref{pres}) with zero pressure and (\ref{denk}) determine
the $R$ dependence of the potentials $V$ and $H$ completely. 
Using Proposition 5  we have chosen the integration constants
($R$ independent functions) as the functions $c_{i}$ ($i=1,2,3$) so that 
$c_{1}=m$ , $c_{2}=e$ and $c_{3}=c$ and

\[
c=c(\tau, a, a_{k}),~~e=e(\tau),
\]
\[
m=\left \{ \begin{array}{ll}
M(\tau)+\epsilon\,\kappa (D-3) e c  & (\mbox{$D \ge 4$}) \label{rel1}\\
M(\tau)-{\kappa \over 2}\,e^2-\epsilon\, \kappa e c  & (\mbox{$D=3$}) 
\label{rel2}
\end{array} 
\right.
\]

\noindent
where $a_{k}$'s are defined in Lemma 1.

\vspace{0.3cm}

\noindent
{\bf Remark 2}.\, The curve $C$ is a geodesic of the curved geometry with
the metric $g_{\mu \nu}=\eta_{\mu \nu}-2V \lambda_{\mu} \lambda_{\nu}$ if and
only if it is a straight line in ${\bf M_{D}}$.
 
\vspace{0.3cm}

\noindent
{\bf Remark 3}.\, When $D=4$, we obtain the Bonnor-Vaidya solutions with
one essential difference. In the Bonnor-Vaidya solutions the parameters
$m$ and $c$ (which are related  through (\ref{rel2}))
depend upon $\tau$ and $a$ only. In our solution, these parameters depend not
only on $\tau$ and $a$ but also on all other scalars $a_{k}$, ($k \ge 1$).

\vspace{0.3cm}

\noindent
{\bf Remark 4}.\, The electromagnetic vector potentials in the flat space
(see Section 2) and in the curved space (see Section 3) are different. It
is known that in four dimensions they are gauge equivalent \cite{bv1}. 
It is this equivalence that led Bonnor and Vaidya to choose the function
$c(\tau, a, a_{k})=-ea$. All other choices of $c$ do not have flat space
limits.
In higher dimensions ($D>4$) such an equivalence 
can not be established. As an example, let 
$$A_{f}^{\mu}={\ddot{z}^{\mu}-a \dot{z}^{\mu} \over R^2}+\epsilon {\dot{z}^{\mu}
\over R^3}$$
be the vector potential (\ref{a66}) in 6-dimensional flat space. 
Let  $A_{\mu}=H \lambda_{\mu}=\Phi_{, \mu}+A_{f\, \mu}$, where
$\Phi$ is the gauge potential to be determined.
Here $A_{\mu}$ is the vector potential (\ref{e51}) in 6-dimensional
curved spacetime.
Except for the static case, it is a simple calculation to show
that there doesn't exist any
$\Phi$ to establish such a gauge equivalence. Hence our higher 
dimensional solutions have no flat space counterparts.
In the static case (the curve $C$ is a straight line)
our solutions are, for all $D$, gauge equivalent to the Tangherlini
\cite{tan} solutions (see also \cite{pery}).
Because of this reason the energy flux expressions in 6-dimensions (see
the next section) in flat and curved spacetimes are different. To obtain
gauge equivalent solutions we conjecture that the Kerr-Schild ansatz has to 
be abandoned.

\vspace{0.3cm}

\noindent
{\bf Remark 5}.\, (a)\, It is easy to prove that $\rho=0$ only when the
curve $C$ is a straight line in ${\bf M_{D}}$ (static case). This means
that there are no accelerated vacuum and electro-vacuum solutions.\\
(b)\, We can have pure fluid solutions when $e=c=0$. In this case 
we have

\begin{eqnarray}
V&=&m R^{3-D}, \nonumber \\
\rho&=&{2-D \over \kappa}\, [a (1-D)M+\dot{M}]\, R^{2-D}
\end{eqnarray}

\noindent
for $D \ge 4$ and 

\begin{eqnarray}
V&=& m, \nonumber \\
\rho&=& {2Ma -\dot{M} \over \kappa R}
\end{eqnarray}

\noindent
for $D=3$. Such solutions are usually called as the {\it Photon Rocket}
solutions \cite{kin}-\cite{ser}. We give here the D dimensional
generalizations of this type of metrics as well.

\section{Radiation due to Acceleration}

In this Section we give a detailed analysis of the energy flux due
to the acceleration of charged sources in the case of the solution
given in Theorem 6. For the  $D > 3$ case, the solution described
by the functions $c$, $e$, and $M$ give the energy density  given in 
(\ref{ro00}).
Remember that at this point $c=c(\tau,a,a_{k})$ and arbitrary. Choosing
$e=$ constant , $c=-e a$ as was done by Bonnor and Vaidya \cite{bv1},
the expression for $\rho$ simplifies and one gets

\begin{eqnarray}
\rho&=&{1 \over \kappa} (2-D) \dot{M} R^{2-D}+{a \over \kappa}(2-D)(1-D)
M R^{2-D} \nonumber \\
&&-{e^2 \over R^2} (\ddot{z}^{\alpha}\, \ddot{z}_{\alpha}+
\epsilon a^2)  
+(3-D)(2-D) e^2 [ \epsilon\,a_{1} R^{2-D} \nonumber \\
&&+ a R^{5-2D}- a R^{1-D}-\epsilon \,D a^2 R^{2-D}]
\label{ro1}
\end{eqnarray}

\noindent
The flux of null fluid energy is then  given by

\begin{equation}
N_{f}=-\int_{S^{D-2}} T_{f}\,^{ \alpha} \, _{\beta} \, \dot{z}_{\alpha}\, 
n^{\beta} R^{D-3}\, d\Omega
\end{equation}

\noindent
and since $T_{f}\,^{\alpha}\,_{\beta}=\rho\, \lambda^{\alpha}\,
\lambda_{\beta}$ for the special case  $p=\Lambda=0$ that we are examining,
one finds that

\begin{equation}
N_{f}=\epsilon \epsilon_{1}\, \int_{S^{D-2}} \, \rho R^{D-2}\, d\Omega
\end{equation}

\noindent
where $\rho$ is given in (\ref{ro1}). The flux of electromagnetic
energy  is similarly given by

\begin{equation}
N_{e}=-\int_{S^{D-2}}\, T_{e}\,^{\alpha}\,_{\beta}\, \dot{z}_{\alpha}\, 
n^{\beta} R^{D-3}\, d\Omega   \label{nel}
\end{equation}

\noindent
and for the solution we are examining, one finds that

\newpage

\begin{eqnarray}
-T_{e}\,^{\alpha}\,_{\beta}\, \dot{z}_{\alpha}\, n^{\beta}&=&
\epsilon_{1} \epsilon\, (3-D)\,e\, [(3-D) e (A-\epsilon) R^{5-2D} \nonumber \\
&&+\dot{e}
 R^{6-2D}+\epsilon \,(c_{,\alpha}\,\dot{z}^{\alpha}) R^{3-D}] 
+\epsilon \epsilon_{1}\,R (c_{,\alpha} c^{, \alpha}) \nonumber \\
&&+(3-D)^2 e^2 (R_{,\alpha}\, n^{\alpha})
R^{4-2D}\nonumber \\
&&+(3-D) e R^{5-2D} \dot{e} (\lambda_{\alpha}\,n^{\alpha}) \nonumber \\
&&+\epsilon\, (3-D) e R^{2-D}\, (c_{,\alpha} n^{\alpha})
\end{eqnarray}

\noindent
for the case $c=c(\tau,a, a_{k})$ and arbitrary. Taking the special case
$e=$ constant, $c=-e a$ of Bonnor-Vaidya \cite{bv1},
this simplifies and hence (\ref{nel}) becomes

\begin{equation}
N_{e}=\epsilon_{1}\, e^2\, \int_{S^{D-2}} d\Omega\, [a^2+\epsilon\,
(\ddot{z}^{\alpha}\,\ddot{z}_{\alpha})]\, R^{D-4}. \label{efl1}
\end{equation}

\noindent
The total energy flux is  given by

\begin{eqnarray}
N&=&N_{e}+N_{f}= \epsilon_{1}\, \int_{S^{D-2}}\,
[ {(2-D)\epsilon \over \kappa}\dot{M}
+{\epsilon \over \kappa} a(2-D)(1-D) M \nonumber \\
&&+(3-D)(2-D) e^2 a_{1} -D(2-D)(3-D) e^2 a^2 ]
\end{eqnarray}

\noindent
for $R$ large enough. For a charge with  acceleration 
$|\ddot{z}_{\alpha}|=\kappa_{1}$,  we have (see the Appendix)

\begin{eqnarray}
&&N={1 \over 2} (2-D)\{ \dot{M} \epsilon +  2(3-D) e^2 (\kappa_{1})^2 \,
\Omega_{D-2}
\nonumber \\
&&-D(3-D) e^2 (\kappa_{1})^2 \, \Omega_{D-3}\,
\Gamma({D-2 \over 2})\, \gamma_{D}\}\epsilon_{1}, \label{fl2}
\end{eqnarray}

\noindent
where

\[
\gamma_{D}=  {-2 \sqrt{\pi} \over \Gamma({D-1 \over 2})}+
2^{D} {\Gamma({D+2 \over 2}) \over \Gamma(D)}. 
\]

\noindent
Notice that when $D=4$, this reduces to the result of Bonnor-Vaidya 
in Ref.\cite{bv1}. We also give the $D=6$ case as another example.

\begin{equation}
N=\left \{ \begin{array}{ll}
\epsilon_{1}\,[-\epsilon \dot{M}-{8\pi \over 3} e^2 (\kappa_{1})^2] & 
(\mbox{$D =4$})\\
\epsilon_{1}\, [-2 \epsilon \dot{M}-{96 \pi^2 \over 15}
e^2 (\kappa_{1})^2]  &(\mbox{$D=6$})
\end{array}
\right. \label{los3}
\end{equation}

\vspace{0.3cm}

\noindent
{\bf Remark 6}.\, To be consistent with Bonnor and Vaidya we take
$\epsilon_{1}=- 1$ which is also consistent with our Remark 1.
Electromagnetic energy flux (\ref{efl1}) is
finite only when $D=4$, but the total energy flux (\ref{fl2}) is finite,
for all $D$ (due to the cancelation of divergent terms in $N_{e}$ and $N_{f}$).

\vspace{0.3cm}

\noindent
{\bf Remark 7}.\, The difference between the energy flux expressions
in the $D=4$ flat space (\ref{los1})  with one obtained
above (\ref{los3}) is due to the scaling factor (${1 \over 4\pi}$).
Apart from this difference the  energy flux expressions obtained
from classical electromagnetism and general relativity are exactly the same
for the special case $c=-ea$ and $M=$ constant. For other choices  we 
obtain different
expressions for the energy flux. For $D > 4$ since we do not have gauge 
equivalent solutions (see Remark 4) the energy flux expressions 
obtained from classical electromagnetism and  general relativity are 
different. 

\vspace{0.3cm}

For the $D=3$ case, substituting the special choice $e=$ constant, $c=-ea$
of \cite{bv1}, the expression for $\rho$ given in (\ref{ro11}) reduces
to

\begin{eqnarray}
\rho&=&-{\dot{M} \over \kappa R}+{2M a \over \kappa R}-{e^2 \over R^2}\, 
(\ddot{z}^{\alpha}\,
\ddot{z}_{\alpha})-\epsilon {e^2 a_{1} \over R}-\epsilon {e^2 a^2 \over R^2}
\nonumber \\
&&-{e^2 a \over R}\, \ln R+ {e^2 a \over R}\,[3\epsilon a
+{1 \over R}-{1 \over 2}].
 \label{ro22}
\end{eqnarray}

\noindent
Similarly the energy flux due to the fluid is  found as

\[
N_{f}=\epsilon \epsilon_{1} \int_{0}^{2\pi}\, \rho \, R\, d \theta
\]

\noindent
where $\rho$ is given in (\ref{ro22}). For the energy flux due to the 
electromagnetic field, one finds

\begin{eqnarray}
-T_{e}\,^{\alpha}\,_{\beta}\, \dot{z}_{\alpha}\, n^{\beta}&=&
\epsilon_{1} \epsilon {e^2 \over R}(A-\epsilon)+
\epsilon \epsilon_{1} e \dot{e} \ln R +
 \epsilon_{1}\, e (c_{,\alpha}\, \dot{z}^{\alpha}) \nonumber \\
&& + \epsilon \epsilon_{1}\,R (c_{,\alpha}\, c^{, \alpha})+{e^2 \over R^2}\, 
(R_{,\alpha}\, n^{\alpha})+
{e \dot{e} \over R} \ln R (\lambda_{\alpha}\,n^{\alpha}) \nonumber \\
&&+\epsilon\,{e \over R} (c_{,\alpha}\, n^{,\alpha})
\end{eqnarray}

\noindent
Here  $c=c(\tau, a , a_{k})$ is an arbitrary function of its arguments. For the 
special case $e=$ constant, $c=-ea$ energy flux due to the electromagnetic
field becomes

\[
N_{e}={e^2 \over R}\,\epsilon_{1}\, \int_{0}^{2\pi}\, 
d\theta\, [a^2+\epsilon (\ddot{z}^{\alpha}\, \ddot{z}_{\alpha})] 
\]

\noindent
The total energy loss is given by

\begin{eqnarray}
N=N_{e}+N_{f}= \epsilon_{1}\, \int_{0}^{2 \pi}\, d\theta \, 
[ - {\epsilon \dot{M} \over \kappa}+
{2 \epsilon M a \over \kappa}- e^2 a_{1}-\epsilon ae^2 \ln R \nonumber  \\
+3 e^2 a^2+{\epsilon \over R} e^2 a-
{\epsilon \over 2} e^2 a]. \label{nnn}
\end{eqnarray}

\noindent
Assuming that the curve $C$ is time like ($\epsilon=-1$),
performing the angular integration in (\ref{nnn}), then taking $R$
large enough we get

\[
 N=[{1 \over 2} \dot{M}+\pi e^2 \kappa_{1}^2 ]\, \epsilon_{1}
\]

\vspace{0.3cm}

\noindent
{\bf Remark 8}.\, The sign of the energy flux expression in three dimensions 
is opposite to the one obtained in other dimensions ($D>3$).

\vspace{0.3cm}

\section{Conclusion}

We have found exact solutions of the $D$-dimensional Einstein Maxwell 
field equations with a null perfect fluid source. Physically these solutions 
describe the electromagnetic and
gravitational fields of a charged particle moving on an arbitrary curve
 $C$ in a $D$-
dimensional manifold. The metric and the electromagnetic vector potential
 arbitrarily
depend on a  scalar, $c(\tau_{0}, a, a_{k})$ which can be related to the 
curvatures
of the curve $C$. In four dimensions with a special choice of this scalar,
 our solution
matches with the Bonnor-Vaidya metric \cite{bv1}. For other choices we 
have different
solutions. In higher dimensions our solutions given in Theorem 6, for all $D$,
can be considered as the accelerated Tangherlini \cite{tan} solutions. 

We have also studied the flux of electromagnetic energy  due to
the acceleration of  charged
particles. We observed that the energy flux  formula, for all dimensions,
depends on the choice of the scalar $c$ in terms of the functions $a$, $a_{k}$
(or the curvature scalars of the curve $C$).
In dimensions $D>4$ electromagnetic and fluid energy fluxes
diverge for large values of $R$ but the total energy flux  is finite. We 
obtained the energy flux expression corresponding to a special choice,
$c=-ea$, for all dimensions.

\vspace{1.5cm}

We would like to thank the referees for their constructive criticisms.
This work is partially supported by the Scientific and Technical Research
Council of Turkey and by the Turkish Academy of Sciences.

\section*{Appendix: Serret-Frenet Frames.}

In this Appendix we shall give the Serret-Frenet frame in four dimensions
which can be easily extended to any arbitrary dimension $D$.
The curve $C$ described in Section 2 has the tangent vector
 $T^{\mu}=\dot{z}^{\mu}$.
Starting from this tangent vector, by repeated differentiation with respect to
the arclength parameter $\tau_{0}$, one can generate an orthonormal frame 
$\{ T^{\mu}, N_{1}^{\mu},N_{2}^{\mu},N_{3}^{\mu} \}$ ,the 
{\it Serret-Frenet frame}.

\begin{eqnarray}
\dot{T}^{\mu}&=&\kappa_{1}\, N_{1}^{\mu},  \\
\dot{N}_{1}^{\mu}&=&\kappa_{1}\,T^{\mu}-\kappa_{2}\, N_{2}^{\mu},  \\
\dot{N}_{2}^{\mu}&=&\kappa_{2}\,N_{1}^{\mu}-\kappa_{3}\,N_{3}^{\mu},\\
\dot{N}_{3}^{\mu}&=&\kappa_{3}\,N_{2}^{\mu}
\end{eqnarray}

\noindent
Here $\kappa_{i},~ (i=1,2,3)$ are the curvatures of the curve $C$ at the point 
$z^{\mu}(\tau_{0})$. The normal vectors $N_{i}, ~(i=1,2,3)$ are space like 
unit vectors.
Hence at the point $z^{\mu}(\tau_{0})$ on the curve we have an orthonormal
 frame which
can be used  as a basis of the tangent space at this point. In Section 2 
we have also defined some scalars 
$$a_{k}={d^k \ddot{z}_{\mu} \over d\tau_{0}^k}\, \lambda^{\mu}$$
where $\lambda_{\mu}=\epsilon T^{\mu}+\epsilon_{1}\,
{n^{\mu} \over R}$. Here $n^{\mu}$ is a space like
vector orthogonal to $T^{\mu}$. Hence we let
$$n^{\mu}=\alpha\, N_{1}^{\mu}+\beta \, N_{2}^{\mu}\, +\gamma\, N_{3}^{\mu}$$
with $\alpha^2+\beta^2+\gamma^2=R^2$. One can choose the spherical angles $\theta \in (0, \pi),~~
\phi \in (0, 2\pi)$ such that
$$\alpha=R \cos \theta,~~ \beta=R \sin \theta \cos \phi,~~\gamma=
R \sin \theta \sin \phi$$
Hence we can calculate the scalars $a_{k}$ in terms of the curvatures and the spherical
coordinates $(\theta, \phi)$ at the point $z^{\mu}(\tau_{0})$. These expressions are
quite useful in the energy flux formulas. As an example we give $a$
and $a_{1}$.

\begin{equation}
a=-\epsilon \epsilon_{1} \kappa_{1}\, \cos \theta,
~~a_{1}=(\kappa_{1})^2-\epsilon \epsilon_{1} \dot{\kappa}_{1}
 \cos \theta+
\epsilon \epsilon_{1} \kappa_{1} \, \kappa_{2} \, \sin \theta \, \cos \phi
\end{equation}

\noindent
All other $a_{k}$'s can be found similarly. Hence for all $k$, these scalars
depend on the curvatures and the spherical angles. 

In four dimensions
the space like vector $n$ is given by (we shall omit the indices
on the vectors)
\[
n=R\,[ N_{1} \, \cos \theta\, + N_{2}\, \sin \theta \cos \phi +
N_{3} \, \sin \theta \sin \phi]
\]
where $\theta \in (0,\pi)$ and $\phi \in (0,2\pi)$.
The line element on $S^2$ is
\[
ds^2=d\theta^2 +sin^2 \theta d\phi^2
\]
The solid angle integral is
\[
\int_{S^2}\, d\Omega=4\pi.
\]
In six dimensions we have

\begin{eqnarray}
n=R\, [ N_{1}\, \cos \theta + N_{2}\, \sin \theta \cos \phi_{1}+
N_{3}\, \sin \theta \sin \phi_{1} \cos \phi_{2} \nonumber \\
+N_{4}\, \sin \theta \sin \phi_{1} \sin \phi_{2} \cos \phi_{3} +
N_{5}\,\sin \theta \sin \phi_{1} \sin \phi_{2} \sin \phi_{3} ]
\end{eqnarray}

\noindent
where $\theta ,\phi_{1}, \phi_{2} \in (0, \pi)$ and $\phi_{3} \in (0,2\pi )$
and
\[
ds^2=d\theta^2+ \sin^2 \theta\, d\phi_{1}^2+\sin^2 \theta \sin^2 \phi_{1}\,
d\phi_{2}^2+\sin^2 \theta \sin^2 \phi_{1} \sin^2 \phi_{2} d\phi_{3}^2.
\]
The solid angle integral is
\[
\int_{S^4} \, d\Omega={8 \pi^2 \over 3}.
\]
The solid angle integral in $D$-dimensions is
\[
\Omega_{D-2} \equiv \int_{S^{D-2}}\, d\Omega={(D-1) \, \pi^{(D-1)/2} \over
\Gamma({D+1 \over 2})}.
\]



\begin{thebibliography}{99}
\bibitem{bv1} W. B. Bonnor and P. C. Vaidya, {\bf General Relativity},
papers in honor of J. L. Synge, Edited by L. O' Raifeartaigh (Dublin
Institute for Advanced Studies) p. 119 (1972).
\bibitem{ks} R. Kerr and A. Schild, {\bf Applications of nonlinear
partial differential equations in mathematical physics}, {\it Proceedings
of Symposia in Applied Mathematics}, (Amer. Math. Soc., Providence, 
R. I., 1965), Vol. XVII, p.199.
\bibitem{gg1} M. G{\" u}rses and F. G{\" u}rsey, {\it J. Math. Phys.},
{\bf 16}, 2385 (1975).
\bibitem{crm} D. Kramer, H. Stephani, M. A. H. Mac Callum, and E. Herlt,
{\bf Exact Solutions of Einstein's Field Equations}, (Cambridge University 
Press, 1980).
\bibitem{rt1} I. Robinson and A. Trautman, {\it Proc. R. Soc. Lon.},
{\bf A 265}, 463 (1962).
\bibitem{new} E. T. Newman, {\it J. Math. Phys}., {\bf 15}, 44 (1974).
\bibitem{rt3} E. T. Newman and T. W. J. Unti, {\it J. Math. Phys}., {\bf 12},
1467 (1963).
\bibitem{hwk} S. W. Hawking and G. F. R. Ellis, {\bf The large scale structure
of space-time}, (Cambridge University Press, Cambridge, 1977).
\bibitem{brt} A. O. Barut, {\bf Electrodynamics and Classical Theory of
Fields and Particles}, (Dover Publications, New York, 1980).
\bibitem{jack} J. D. Jackson, {\bf Classical Electrodynamics}, (John Wiley
and Sons, New York, 1975).
\bibitem{spv} M. Spivak, {\bf A Comprehensive Introduction to Differential 
Geometry }, (Publish or Perish Inc., 1979).
\bibitem{guv} G. Arrega, R. Capovilla, and J. G{\" u}ven, {\it Class. 
Quantum Grav.},
{\bf 18}, 5065 (2001). 
\bibitem{go1} M. G{\" u}rses and {\" O}. Sar{\i}o\u{g}lu, {\bf Radiation due to
acceleration in $D$-Dimensional Kerr-Schild Geometry}, (in preparation).
\bibitem{go2} M. G{\" u}rses and {\" O}. Sar{\i}o\u{g}lu, {\bf 
Accelerated Born-Infeld Metrics In Kerr-Schild Geometry}, 
submitted for publication.
\bibitem{tan} F. R. Tangherlini, {\it Nuovo Cimento}, {\bf 77}, 636 (1963).
\bibitem{pery} R. C. Myers and M. J. Perry, {\it Annals of Physics},
{\bf 172}, 304 (1986).
\bibitem{kin} W. Kinnersley, {\it Phys. Rev}., {\bf 186}, 1353 (1969).
\bibitem{bon} W.B. Bonnor, {\it Class. Quantum Grav}. {\bf 11}, 2007 (1994).
\bibitem{dam} T. Damour, {\it Class. Quantum Grav}. {\bf 12}, 725, (1995).
\bibitem{ser} S. Dain, O. M. Moreschi, and R. J. Gleiser, {\bf Photon Rockets
and the Robinson Trautman Geometries}, {\tt arXiv:gr-qc/0203064}.

\end{thebibliography}
\end{document}